\documentclass[journal]{IEEEtran}
\usepackage[utf8x]{inputenc}

\usepackage{amscd,amsfonts,amssymb}
\usepackage{amsmath}
\usepackage{xspace}
\usepackage{color}
\usepackage{import}
\usepackage{algorithmicx}
\usepackage[table]{xcolor}
\usepackage{pgfplots}
\usepackage{tikz,ifthen}
\usetikzlibrary{spy} 
\usetikzlibrary{patterns}
\usetikzlibrary{dsp,chains}
\usepackage{timath}
\usepackage{mathtools}
\usepackage[font=small,labelfont=bf]{caption}
\usepackage{subcaption}
\usepackage{multirow}
\usepackage{xstring}
\usepackage{mytikz} 
\usepackage{outlines}
\usepackage{accents}  
\usepackage{algpseudocode}
\usepackage{algorithm}
\usepackage{graphicx}

\usepackage{subcaption}

\usepackage{todonotes}

\usepackage{makeidx}
\makeindex

\usepackage[unicode=true,hidelinks,linktoc=all]{hyperref}
\hypersetup{nolinks=true}  

\usepackage{etoolbox}
\newtoggle{finalorsubmission}

\usepackage[printonlyused]{acronym}
\makeatletter
\patchcmd{\AC@@acro}{] #3}{] \MakeUppercase #3}{}{}
\patchcmd{\AC@@acro}{] #3}{] \MakeUppercase #3}{}{}
\makeatother

\usepackage{url}
\usepackage[super]{nth}

\usepackage[nomain, acronym, toc, shortcuts]{glossaries}

\newacronym{2d}{2D}{\emph{two dimensional}}
\newacronym{3d}{3D}{\emph{three dimensional}}
\newacronym{5g}{5G}{\emph{fifth generation}}
\newacronym{admm}{ADMM}{\emph{alternating direction method of multipliers}}
\newacronym{ahrs}{AHRS}{\emph{attitude and heading reference system}}
\newacronym{aoa}{AoA}{\emph{angle of arrival}}
\newacronym{ap}{AP}{\emph{Access Point}}
\newacronym{bab}{B\&B}{\emph{branch and bound}}
\newacronym{blu}{BLU}{\emph{best linear unbiased}}
\newacronym{blue}{BLUE}{\emph{best linear unbiased estimator}}
\newacronym{bp}{BP}{\emph{Basis Pursuit}}
\newacronym{bpdn}{BPDN}{\emph{Basis Pursuit Denoising Sensing}}
\newacronym{cu}{CU}{\emph{central unit}}
\newacronym{dcu}{DCU}{\emph{data collection unit}}
\newacronym{ds}{DS}{\emph{Dantzig Selector}}
\newacronym{cdf}{cdf}{\emph{cumulative distribution function}}
\newacronym{crb}{CRB}{\emph{Cramer-Rao Bound}}
\newacronym{cs}{CS}{\emph{compressed sensing}}
\newacronym{csi}{CSI}{\emph{channel state information}}
\newacronym{dtdoa}{DTDoA}{\emph{differential TDoA}}
\newacronym{dtoa}{DToA}{\emph{differential ToA}}
\newacronym{ehf}{EHF}{\emph{extremely high frequency}}
\newacronym{esprit}{ESPRIT}{\emph{estimation of signal parameters using rational invariance}}
\newacronym{fc}{FC}{\emph{fusion center}}
\newacronym{gp}{GP}{\emph{grid point}}
\newacronym{gps}{GPS}{\emph{global positioning system}}
\newacronym{ict}{ICT}{\emph{information and communication technology}}
\newacronym{iid}{iid}{\emph{identically and independently distributed}}
\newacronym{ilp}{ILP}{\emph{integer linear program}}
\newacronym{imu}{IMU}{\emph{inertial measurement unit}}
\newacronym{iot}{IoT}{\emph{internet of things}}
\newacronym{ism}{ISM}{\emph{industrial, scientific and medical}}
\newacronym{kkt}{KKT}{\emph{Karush-Kuhn-Tucker}}
\newacronym{knn}{KNN}{\emph{K-Nearest Neighbor}}
\newacronym{ln}{$\mathcal{LN}$}{\emph{log-normal}}
\newacronym{los}{LoS}{\emph{line of sight}}
\newacronym{lpwan}{LPWAN}{\emph{low power wide area network}}
\newacronym{lrwpan}{LR-WPAN}{\emph{low rate wireless personal access network}}
\newacronym{lte}{LTE}{\emph{long-term evolution}}
\newacronym{nlos}{NLoS}{\emph{non line of sight}}
\newacronym{mac}{MAC}{\emph{medium access control}}
\newacronym{milp}{MILP}{\emph{mixed integer linear program}}
\newacronym{mip}{MIP}{\emph{mixed integer program}}
\newacronym{miqp}{MIQP}{\emph{mixed integer quadratic program}}
\newacronym{ml}{ML}{\emph{maximum likelihood}} 
\newacronym{mle}{MLE}{\emph{maximum likelihood estimator}} 
\newacronym{mlp}{MLP}{\emph{multilayer perceptron}}
\newacronym{mmw}{mmW}{\emph{millimeter wave}} 
\newacronym{mse}{MSE}{\emph{mean square error}}
\newacronym{mu}{MU}{\emph{measurement unit}}
\newacronym{music}{MUSIC}{\emph{multiple signal classification algorithm}}
\newacronym{mvu}{MVU}{\emph{minimum variance unbiased}}
\newacronym{mvue}{MVUE}{\emph{minimum variance unbiased estimator}}
\newacronym{nsp}{NSP}{\emph{null space property}}
\newacronym{pdf}{pdf}{\emph{probability distribution function}}
\newacronym{pdoa}{PDoA}{\emph{power difference of arrival}}
\newacronym{ple}{PLE}{\emph{path-loss exponent}}
\newacronym{plf}{PLF}{\emph{path-loss factor}}
\newacronym{pmf}{pmf}{\emph{probability mass function}}
\newacronym{pmu}{PMU}{\emph{power measurement unit}}
\newacronym{prmse}{PRMSE}{\emph{positioning root mean square error}}
\newacronym{qp}{QP}{\emph{quadratic program}}
\newacronym{rip}{RIP}{\emph{restricted isometry property}}
\newacronym{rf}{RF}{\emph{radio-frequency}}
\newacronym{rfid}{RFID}{\emph{radio-frequency identification}}
\newacronym{rmse}{RMSE}{\emph{root mean square error}}
\newacronym{rss}{RSS}{\emph{received signal strength}}
\newacronym{rv}{\emph{r.v}}{\emph{random variable}}
\newacronym{ser}{SER}{\emph{symbol error rate}}
\newacronym{shf}{SHF}{\emph{super high frequency}}
\newacronym{sln}{$\mathcal{SLN}$}{\emph{sum log-normal}}
\newacronym{sm}{SM}{\emph{smart meter}}
\newacronym{sn}{SN}{\emph{sensor node}}
\newacronym{snr}{SNR}{\emph{signal-to-noise ratio}}
\newacronym{svm}{SVM}{\emph{support vector machine}}
\newacronym{tdm}{TDM}{\emph{time division multiplexing}}
\newacronym{tdma}{TDMA}{\emph{time division multiplexing access}}
\newacronym{tdoa}{TDoA}{\emph{time difference of arrival}}
\newacronym{tee}{TEE}{\emph{total estimation error}}
\newacronym{toa}{ToA}{\emph{time of arrival}}
\newacronym{tof}{ToF}{\emph{time of flight}}
\newacronym{uwb}{UWB}{\emph{ultra wideband}}
\newacronym{vf}{VF}{\emph{virtual fingerprinting}}
\newacronym{vlc}{VLC}{\emph{visible light communication}}
\newacronym{wpan}{WPAN}{\emph{wireless personal access network}}
\newacronym{wsn}{WSN}{\emph{wireless sensor network}}

\makeglossaries   
\makeindex
\makeatletter

\def\capital#1{%
	\let\tc@w\@empty
	\protected@edef\tmp{\noexpand\tc@a#1\relax}\expandafter\tc@uc@\tmp}

\def\tc@a{\futurelet\tmp\tc@aa}

\def\tc@aa{%
	\ifcat a\noexpand\tmp\expandafter\tc@ab
	\else\expandafter\tc@ac\fi}

\def\tc@ab#1{\edef\tc@w{\tc@w#1}\tc@a}

\def\tc@ac{%
	\csname tc@@\tc@w\endcsname\expandafter\tc@uc\tc@w
	\let\tc@w\@empty
	\ifx\tmp\@sptoken\let\next\tc@sp
	\else\ifx\tmp\relax\let\next\relax
		\else\let\next\tc@nxt
		\fi\fi\next}

\def\tc@sp#1{ \tc@a#1}
\def\tc@nxt#1{#1\tc@a}

\def\tc@uc#1{\uppercase{#1}}
\def\tc@uc@#1#2{\uppercase{#1#2}}

\let\tc@@the\@gobbletwo
\let\tc@@and\@gobbletwo
\let\tc@@in\@gobbletwo
\let\tc@@using\@gobbletwo
\let\tc@@a\@gobbletwo
\let\tc@@with\@gobbletwo
\let\tc@@under\@gobbletwo
\let\tc@@for\@gobbletwo
\let\tc@@of\@gobbletwo
\let\tc@@via\@gobbletwo
\let\tc@@at\@gobbletwo
\let\tc@@to\@gobbletwo

\makeatother
\usepackage[thmmarks]{ntheorem}
\theoremheaderfont{\bfseries}
\theorembodyfont{\normalfont}
\theoremseparator{:}

\newtheorem{remark}{Remark}

\theoremsymbol{$\blacksquare$}   

\usepackage[noabbrev]{cleveref}
\crefformat{equation}{#2(#1)#3}
\crefformat{figure}{Fig. #2#1#3}
\crefformat{remark}{Remark #2#1#3}
\crefformat{section}{Sec. #2#1#3}
\crefformat{table}{Tab. #2#1#3}
\crefformat{algorithm}{Alg. #2#1#3}

\usepackage{rotating}
\usepackage{tablefootnote}  

\usepackage{xinttools} 
\usepackage[shortlabels]{enumitem} 
\usepackage{marvosym} 
\usepackage[numbers,sort&compress,square]{natbib}

\newcommand{\subparagraph}{}
\usepackage{titlesec}
\titlespacing*{\section}
{0pt}{2.5ex plus 1ex minus .2ex}{2.3ex plus .2ex}
\titlespacing*{\subsection}
{0pt}{1.5ex plus 1ex minus .2ex}{1.3ex plus .2ex}

\allowdisplaybreaks

\newcommand{\defterm}[1]{\noindent \textbf{#1} \enspace}

\newcommand{\set}[1]{\mathbb{#1}}
\NewDocumentCommand \idx { o } {\IfNoValueTF {#1} {\set I} {\setI_#1} }
\newcommand{\grid}{\mathcal G}
\newcommand{\power}{\mathcal P}

\DeclareRobustCommand{\pmin}{\underaccent{\bar}{P}}
\DeclareRobustCommand{\pmax}{\bar P}
\DeclareRobustCommand{\amin}{\underaccent{\bar}{\alpha}}
\DeclareRobustCommand{\amax}{\bar \alpha}
\DeclareRobustCommand{\bmin}{\underaccent{\bar}{\beta}}
\DeclareRobustCommand{\bmax}{\bar \beta}
\DeclareRobustCommand{\nmin}{\underaccent{\bar}{N}}
\DeclareRobustCommand{\nmax}{\bar N}

\newcommand{\nphard}{$\mathcal{NP}$-hard\xspace}
\newcommand{\sn}{\check}      
\newcommand{\gp}{\tilde}        
\newcommand{\est}{\hat}     
\newcommand{\lerr}{\delta}     
\newcommand{\perr}{\rho}     

\newcommand{\lrmse}{\overline \lerr}     
\newcommand{\prmse}{\overline \perr}     
\newcommand{\armse}{\overline \alpha}     
\newcommand{\brmse}{\overline \beta}     

\newcommand{\muk}{\mu_k}      
\newcommand{\sigk}{\sigma_k}  

\def\ord#1{$#1^\text{th}$\xspace}
\def\d#1{\text{d}#1}

\DeclareMathAlphabet{\mathpzc}{OT1}{pzc}{m}{it}
\DeclareMathOperator{\EXPOP}{\mathcal E}

\DeclareMathOperator{\bigo}{\mathcal O}

\DeclareMathOperator{\gauss}{\mathcal N}

\newcommand{\nocontentsline}[3]{}
\newcommand{\tocless}[2]{\bgroup\let\addcontentsline=\nocontentsline#1{#2}\egroup}

	\def\ts{\pgfuseplotmark{triangle*}} 
	\def\ss{\pgfuseplotmark{star}} 
	\def\markersize{5 pt} 

\newcommand{\st}{\ensuremath{\operatorname{s.t.}\hspace{0.2cm}&}}
\newcommand{\minimize}[1]{\ensuremath{\underset{#1}{\min}\hspace{0.2cm}&}}

\newcommand{\sensmat}{\boldsymbol{\Phi}}
\newcommand{\orthmat}{\boldsymbol{\Psi}}
\newcommand{\csmat}{\bQ}
\newcommand{\obssym}{\phi}
\newcommand{\orthsym}{\psi}
\newcommand{\cssym}{q}

\newcommand{\lone}{$\ell_1$\xspace}
\newcommand{\lonenorm}{$\ell_1$-norm\xspace}
\newcommand{\lzeronorm}{$\ell_0$-norm\xspace}

\newcommand{\loneloc}{$\ell_1$-localization\xspace}

\newcommand{\normp}[2]{\ensuremath{\lVert #1 \rVert_{_{#2}}}}
\newcommand{\normt}[1]{\normp{#1}{2}}

\newcommand{\normz}[1]{\normp{#1}{0}}

\newcommand{\opt}[1]{\ensuremath{#1^\star}}

\newcommand\setSmax{\ensuremath{\overline{\idx}_{M_0}}}

\newcommand\parti{\ensuremath{\boldsymbol \Pi}}
\newcommand\partSmax{\ensuremath{\overline{\parti}_{M_0}}}
\newcommand\kmean{k-means clustering\xspace}
\newcommand{\knn}{$k$-NN\xspace}

\newcommand{\dxm}{\d \gp x_m}
\newcommand{\dym}{\d \gp y_m}
\newcommand{\dpm}{\d \gp p_m}
\newcommand{\da}{\d \alpha}
\newcommand{\db}{\d \beta}

\newcommand\ple{\gls{ple}\xspace}
\newcommand\plf{\gls{plf}\xspace}
\newcommand\rss{\gls{rss}\xspace}

\newcommand{\linfk}{\bar{f}_k}

\title{Localizing an Unknown Number of mmW Transmitters Under Path Loss Model Uncertainties}

\author{\IEEEauthorblockN{Ehsan~Zandi, and Rudolf~Mathar}\\
Institute for Theoretical Information Technology, RWTH Aachen University, Germany\\
Email: \{zandi,~mathar\}@ti.rwth-aachen.de}

\begin{document}
\maketitle
\begin{abstract}
This work estimates the position and the transmit power of multiple co-channel wireless transmitters under
model uncertainties. The model uncertainties include the number of the targets and the parameters
of the path-loss model which enable the system to cope with changes in the weather conditions and in mmW ranges.
The problem is solved by an unbiased estimator. The underlying
complicated optimization problem has a combinatorial nature
that selects the best grid points as the location of the targets. 
The combinatorial problem is converted
to a convex form by means of \lone-regularization, which enables
locating off-grid targets.
Simulations show that the proposed algorithm solves the problem with very high accuracy in the
absence of noise and shadowing.

\end{abstract}


\section{Introduction} \label{introduction}
It is envisaged that the majority of  applications in the context of the internet of things and 5G mobile networks depend on the
location awareness to deliver better services. 
This works studies a \gls{rss}-based technique since it is simpler to implement and has lower costs, compared to 
the \gls{tdoa} or \gls{aoa}. Despite its lower positioning accuracy,
\gls{rss} localization is beneficial in case precision can be somewhat compromised for price.
The main challenge in \gls{rss}-based localization is the uncertainties about the path-loss model parameters \cite{patwari2005}.
This issue is addressed in this work.
The \gls{rss}-based localization for a single target with unknown transmit power is studied in many publications, such as \cite{lee2009},
where the transmit power cancels out upon dividing the \gls{rss} of two different receivers. The remaining of the problem is a standard 
multilateration problem. This technique is known as differential or ratio of \gls{rss} and is not applicable to
multi co-channel targets.

\subsection{Path Loss Model Uncertainties}
\label{sec:modeluncert_ple}
In the free space communication,
$\alpha=2$ is a good approximate for the \ple. However, in other types of environments $\alpha$ is different from 2.
Indeed, its value  not only depends on the propagation environment but also changes
due to seasonal or weather conditions \cite{rappaport_book}. Furthermore, the free
space model does not accurately describe the path loss in indoor environments. Numerous path loss
models have been proposed based on measurements in different cities and locations \cite[Sec. 3.11]{rappaport_book}.
Furthermore, the estimation of the \gls{ple} has been the research topic of several works in the literature.

For instance, \cite{mao2007}
exploits the probabilistic information about the distances between nodes and tries to estimate $\alpha$. It also
applies the technique of \emph{distance geometry problem} if such probabilistic information is not available. 
However, it works only for one target with known transmit power. The advantage of such an
idea is that it works with only \gls{rss} measurements and avoids distance calibrating. 
Calibrating based on distance measurements is costly, difficult, and even in some environments impossible
since the distances between transmitter and receivers should be physically measured.  
Such a calibration is also prone to the changes in the values of $\alpha$, which necessitates a new round of calibration.
The existing works in the literature of localization can be classified to the following cases:
\begin{enumerate}
\item Both transmit power and \gls{ple} are known, e.g., \cite{li_TVT_2007}.
\item Transmit power is unknown, \gls{ple} is known, e.g., \cite{kong2008}.
\item \Gls{ple} is known, transmit power is unknown, e.g., \cite{li_TWC_2006}.
\item Neither transmit power nor \gls{ple} is known, e.g., \cite{chan2012ccece}.
\end{enumerate}
Nevertheless, none of the existing works deals with the problem of multi-target localization
in case of unknown \gls{ple}.
In general, the severity of signal strength attenuation depends on a variety of factors, such as wavelength, 
gain and directivity of antennae,
obstacles and big objects in the propagation environment, the height of antennae,
and the existence of the \gls{los} \cite{rappaport_book}. 
This also includes the shape and type of buildings and walls. 
Therefore, the Friis formula is not accurate in practice.
Conventionally, a common way for network planning is using empirical models, which
adds some correction terms to the basic path loss formula. One famous example of 
such a model is the Hata model \cite{hata80}, where the effect of
antennae and other correction factors are included:
\begin{equation}
P_r\text{[dB]}=P_t\text{[dB]}-(44.9-6.55\log h_t)\log d+P\,,
\end{equation}
where $P$ depends on the carrier frequency, the height $h_t$ of transmit antenna as well as 
the height of receive antenna \cite[Eq. 3.82]{rappaport_book}.
This model is analogous to the free space path loss equation,
by letting \gls{ple} be equal to $\alpha=4.49-0.655\log h_t$.
Another conventional method of predicting the path loss is
using an average (effective) value of $\alpha$ acquired by
extensive radio filed measurements, which leads to an empirical path loss model. This
approach is well-known in indoor scenarios, where the effect of walls and their types, number
of floors or windows et cetera on path loss cannot be neglected. For instance, the model in 
\cite[Eq. 3.94-3.95]{rappaport_book}:
\begin{equation}
\overline{PL}(d)\text{[dB]}=\overline{PL}(d_0)\text{[dB]}+10\alpha\log(\frac{d}{d_0})\,,
\end{equation}
where $d_0$ is the reference distance and $\alpha$ varies between 1.81 to 5.22 in different 
locations. Depending on the number of measurement points and the size of the area, over which the measurements
are averaged, predicting path loss using this model leads to different
amounts of deviations from the actual value of path loss. In order to account for such uncertainty,
an additive Gaussian zero-mean \gls{rv} $X_\sigma$, with the variance of $\sigma^2$, can be added to the formula:
\begin{equation}
PL(d)\text{[dB]}=PL(d_0)\text{[dB]}+10\alpha\log(\frac{d}{d_0})+X_\sigma\,,
\end{equation}
where $\alpha$ is an average value (slope of a linear curve fitting over measurement points) in a particular environment. Its 
value is usually higher than 2 in urban areas.
Moreover,  the actual value of path loss and thus $\alpha$ is subject to change. For instance, in an indoor scenario
closing or opening windows may change the path loss level.

\subsection{Rain Fade in mmW}
\label{sec:modeluncert_rainfade}
The water molecules in any form of precipitation, e.g.,  rain, fog, humidity, or snow, intensify
the attenuation of the electromagnetic waves \cite[Ch. 7]{clark_book}. 
Precipitations are the main source of changes in the amount of path loss. The \emph{international telecommunication union 
radiocommunication sector} (ITU-R) has provided methods in the \emph{recommendation P.530} to calculate the excessive path loss
imposed by precipitations.
According to P.530 in addition to the conventional propagation effects such as fading and diffraction, the following
reasons attenuate the signal, excessively:
\begin{itemize}
\item Absorption by the precipitations such as rain or snow.
\item Atmospheric gases that absorb signal, i.e., \emph{dielectric}.
\end{itemize}
In the frequencies over 10GHz, the rain absorption (rain fade) is severe as the size of
the raindrops is comparable with the wavelength of signals. This effect becomes more severe in the case
of mmW, over 30GHz, where the absorption by atmospheric gases and humidity 
as well as rain fade become more challenging. 
Based on ITU-R recommendation PN.837 such an excessive path loss in frequencies up to
40GHz and in \gls{los} situation is given by
\begin{equation}
\gamma_r(\frac{\text{dB}}{\text{m}})=a_0R^{b_0},\,
\end{equation}
where $R$ is the rate of rainfall in mm/hr exceeded for $0.01\%$ of an average year (annual statistic), also
$a_0$ and $b_0$ depend on local climate conditions. The recommended values of
$R$ are provided by PN.837 for different climate zones. They can also be acquired from weather
monitoring centers. 
In the literature there exist several publications that characterize
the rain fade attenuation in different areas and climate zones, e.g., \cite{kestwal2014}.

In summary, there are a few points to highlight:
\begin{enumerate}
\item \gls{rss} (in dB) decreases not only with a constant rate of $10\alpha$ (per decade) but also 
with an additional linear term $\gamma_r$ (per m), as the distance between the transmitter
and receiver increases.
\item $\alpha$ is not always 2, but it lies into a range  between $\amin=1$ to $\amax=6$. The values below 2
are relevant to the kinds of propagation channel which behave like a waveguide, such
as tunnels. Contrarily, $\amax$ represents environments with very harsh shadowing effects.
\item $\gamma_r\geq 0$ depends on many factors including the rainfall rate. Thus, it changes
over time.  More importantly, the values recommended by ITU-R or the estimated values thereupon
can be inaccurate, which degrades the localization accuracy.
\end{enumerate}

\subsection{Unknown Number of Targets}
\label{sec:modeluncert_targnum}
Regarding uncertainties, the most challenging parameter to estimate is
the number of targets since they are non-cooperative.

\subsection{The Contribution}
This work assumes a log-normal shadowing path-loss model, where
multiple co-channel transmitters cause interference on one another. Thus,
multilateration technique, unlike the single target case, is impossible.
To the best of knowledge of the authors,
there is no work with similar assumptions, except for 
\cite{zandi_iswcs2018,zandi_wisee2018,zandi_wsa2019}, where, unlike this work, the number of
targets and the \gls{ple} are assumed to be known.
Furthermore, in this work, the path loss model is suitable for
mmW ranges in different weather conditions.
Since no work with similar assumptions and system model has been found, the results of
this paper could not be compared with any other work, unfortunately.

The organization of this paper is as follows: the system model is described in \cref{sec:system_model},
while \cref{sec:l1local} presents the statistical properties of the \gls{rss} and
the proposed \loneloc technique.
The performance of the presented
technique is evaluated by computer simulations in \cref{sec:simulation}.

{\small {\bf Notations:}
\Cref{tab:notations} shows all mathematical notations of this paper.}

\renewcommand{\arraystretch}{1.3}
\begin{table}[b] 
\caption{Summary of general mathematical notations}\label{tab:notations}
\begin{center}
\begin{tabular}{ c|l }
\textbf{Notation} & \textbf{Description} \\
 \hline
$\setN$ & set of all integer positive and non-zero numbers \\
$\setR$  & set of all real  numbers \\
$\bmx$ & column vector $\bmx$ with entries $x_i$\\
$\bmx'$  & transpose of  vector $\bmx$  \\
$\bX$ & matrix $\bX$ with entries $x_{ij}$ or $[\bX]_{ij}$\\
$\normz{\bmx}$ & \lzeronorm, i.e., the number of non-zero entries of $\bmx$  \\
%
%
$\opt{(\cdot)}$                 & optimal solution of an optimization problem \\
$\est{x}$                 & estimation of the unknown variable $x$ \\
\end{tabular}
\end{center}
\end{table}

%

\section{System Model} \label{sec:system_model}
The system of consideration consists of $N\in\setN$ active targets with unknown positions
and $K\in\setN$ passive \glspl{sn} with known positions. Each target 
transmits a signal with the unknown power $\pmin \leq p_n \leq \pmax$,
where $\pmin,~\pmax\geq 0$ are, the lowest and highest possible values for the transmit power.

The propagation channel is based on the log-normal shadowing attenuation model presented in \cite{rappaport_book}.
In a multi-source scenario, the \gls{rss} $r_k$ at sensor $k$ is the sum of different terms corresponding to 
different target signals \cite{cardieri_rappaport2000,stuber_book}:
\glsreset{ple}
\glsreset{plf}
\begin{align}
\label{eq:modeluncert_ple_basic_formula}
r_k=\sum\limits_{n\in\idx_N} p_n d_{kn}^{-\alpha} \beta^{d_{kn}}\,10^\frac{\zeta_{kn}}{10}\,,
\end{align}
where $d_{kn}$ is the distance between sensor $k$ and \ord n target, $\alpha$ is the path-loss exponent.
$\zeta_{kn}\sim\gauss(0,\sigma^2_{kn})$ is a zero-mean 
Gaussian random variable with the power of $\sigma^2_{kn}$ that models the log-normal shadowing 
and is assumed to be \gls{iid}.
The parameter $\beta=10^\frac{-\gamma_r}{10}$ represents the rain fade.
        \begin{center}
            \fbox{\begin{minipage}{23.1em}
In this work, $\beta$ is termed \plf analogous to the \ple. To be precise, it points out the fact that $\beta$ appears as a 
multiplicative term, and not as an exponent, in the path loss formula \cref{eq:modeluncert_ple_basic_formula}.
            \end{minipage}}
        \end{center}
The coefficient $c_0$ depends on many factors such as
the gains of antennae and the wavelength. Without loss of generality and for the sake of simplicity, it is assumed that $c_0=1$. 

The thermal additive noise is neglected since shadowing has a much stronger effect on \gls{rss} 
compared to the thermal noise \cite{li_TVT_2007,chan2016milcom}.  
Besides, the effect of additive noise can be somewhat compensated using methods of blind estimation of the noise power,
e.g., \cite{gerkmann2012}.
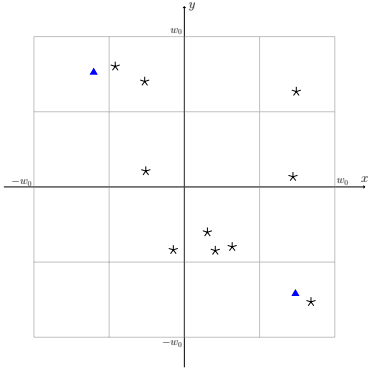
\begin{figure}
	\normalsize
	\centering
	\scalebox{0.2}{
\usetikzlibrary{calc}
\tikzset{label/.style={draw=black, thick,rounded corners=.25ex, text width=10mm, text badly centered,  inner sep=1ex,  minimum height=25mm}}
\begin{tikzpicture}
	\def\K{10} 
	\def\N{2}  
	\def\w{10} 
	\def\u{1}  
  \def\G{5}  
	\def\ss{\pgfuseplotmark{star}}      
	\def\ts{\pgfuseplotmark{triangle*}} 
  \pgfmathsetmacro\gstep{2*\w/(\G-1)*\u}  
	\pgfmathsetseed{8}   
	\def\tn{\node[mark size=7pt,color=blue]  at (rand*\w,rand*\w) {\ts};}
	\def\sn{\node[ultra thick,mark size=9pt,color=black] at (rand*\w,rand*\w) {\ss};}

  \draw[thick,->] (0,-\w-2) -- (0,\w+2) node[anchor=west,xshift=5pt] {\Huge$y$};
	\draw[thick,->] (-\w-2,0) -- (\w+2,0)node[anchor=south,yshift=5pt] {\Huge$x$};
	\foreach\x in {1,...,\N}
		{
			\tn
		}
	\foreach\x in {1,...,\K}
		{
			\sn
		}

	\draw[step=\gstep,gray,ultra thin,opacity=0.5] (-\w,-\w) grid (\w,\w);
  \node[xshift=15pt,yshift=10pt] at (\w,0) {\huge$w_0$};
  \node[xshift=-23pt,yshift=10pt] at (-\w,0) {\huge$-w_0$};
  \node[xshift=-15pt,yshift=10pt] at (0,\w) {\huge$w_0$};
  \node[xshift=-23pt,yshift=-10pt] at (0,-\w) {\huge$-w_0$};
\end{tikzpicture}}\caption{A wireless sensor network with $K=10$ 
		sensors (		\protect\tikz \protect\node[semithick,mark size=2pt,color=black] at (0,0) {\protect\ss}; )
		and $N=2$ targets ( \protect\tikz \protect\node[mark size=2pt,color=blue] at (0,0) {\protect\ts}; )
		. The grid granularity is $G=5$.}
	\label{fig:system_block_diagram}
\end{figure}

\begin{remark}
The position estimation becomes
very challenging if the number of targets is unknown.
In what follows, the variable $\nu$ is introduced to estimate the actual number of targets.
It is assumed $\nu$ is bounded between a minimum possible number $\nmin\in\setN$ 
and a maximum  number $\nmax\in\setN,\,\nmax\geq \nmin$, that is
\begin{align}
\nu\in\{\nmin,\cdots,\nmax\}\,.
\end{align}
Note in case the number of targets is known $N=\nmin=\nmax$. Otherwise, 
the choice of $\nmin=1$ is reasonable since there should exist at least one active target.
\end{remark}

\begin{remark}
In the this work, the actual values of $\beta$, $\alpha$, and $\nu$ are unknown.
It is further assumed that $\amin\leq \alpha\leq \amax$ and $\bmin<\beta\leq \bmax$.
While $\bmax=1$ corresponds to no rain fade conditions or frequencies below 3GHz, $\bmin$ is the minimum possible value of \plf,
corresponding to the highest possible rain attenuation. According to ITU-R recommendations, the rain fade amounts to
0.035-0.04 dB/m for frequencies over 100 GHz in very heavy rain, i.e., over 120 mm/hr, and typhoon situations.
Therefore, the value $\bmin=0.96\leq 10^{-0.004}$ is considered, in this work, to be a lower bound of $\beta$.  
This is a loose lower bound since it corresponds to extreme weather conditions in 
mmW ranges, i.e., over 30GHz.

Considering the rain fade in localization scenarios, in single target case, has recently received a significant attention.
One important application of such a scenario
is the \emph{find and rescue} operation in  emergency-related situations in extreme weather conditions.
For instance, \cite{beritelli2018} estimates the rainfall intensity based on the received signal
at a 4G mobile node using neural networks.
Works \cite{fang2016,fang2018} consider the rain fade
in GSM-1.8GHz for distance estimation (not the position) of a single target. In their scenario,
only one receiver is needed since the \ple, \plf and transmit power are assumed to be known.
The authors solve the problem using Newton-Raphson method and failed to see
that the solution has the form of the \emph{Lambert W} function,
which exists in closed-form. The closed-form solution is however shown
in \cite {hosseini2010} in the context of underwater communication. 
Nevertheless, without assuming the
transmit power is known, it becomes impossible to find a closed-form solution
to this problem. The problem becomes even harder if the values for
$\alpha$ and $\beta$ are to be estimated. The current work intends to do position estimation
for the \emph{multi-target} scenario, given the assumption that $\alpha$, $\beta$, transmit
power of targets (different from one another), and most importantly the number of
targets are unknown.
\end{remark}
\begin{remark}
The path loss  model \cref{eq:modeluncert_ple_basic_formula} does not apply only to the rain fade and
the frequencies over 10GHz. A similar path loss model has been also 
proposed by Devasirvatham in \cite{devasirvatham90}
  for indoor multi-floor buildings, where the path loss follows
  the free space model, i.e., $\alpha=2$ plus an additional linear term (dB/m),
  please see \cite[Eq. 3.96]{rappaport_book}.
\end{remark}
The area of observation is assumed to be a square in the range of $[-w,w]$, $w\geq 0$ 
in both x- and y- axes, in the Cartesian coordinate system. The targets and sensors are 
randomly distributed within the area. The ordered pair $(\sn x_k,\sn y_k)$ stands
for the coordinate of \ord k sensor node, while target $n$ is located at the unknown position $(x_n,y_n)$.
Assuming that the fusion center acquires the values of \gls{rss} $r_k$ 
of the \ord k sensor 
error-freely upon successful communication from \gls{sn}, it has to solve the following system of nonlinear equations
\begin{align}
\label{eq:rss_true_targets}
  r_k=\sum_{n\in\idx_N} \frac{p_n\beta^{^{\sqrt{(x_n - \sn x_k)^2+(y_n - \sn y_k)^2}}}}{\left(\sqrt{(x_n - \sn x_k)^2+(y_n - \sn y_k)^2}\right)^{\alpha}}\ ,
\end{align}
to estimate $N$, $\alpha$, $\beta$, the position $(x_n,y_n)$, and the transmit power $p_n$ of each of the $N$ targets.

Such a system of equations is extremely hard to solve. It is, nevertheless,
solved in this work by a low-complexity heuristic. 
First, the area needs to be discretized into a grid of 
granularity of $G\in\setN$ which means $G^2$ grid points. Let $\grid^w_G(x,y)$
be the grid set centered at the point $(x,y)$ of width $2w\geq 0$ and the
granularity $G$, then $\grid^w_G(x,y)$ is defined by
\begin{align}
  \big\{(x-w+&(i-1)\Delta_g,y-w+(j-1)\Delta_g)\,|\,i,j\in\idx_G\big\}\,,\label{eq:local_rss_grid_set}
\end{align}
where $\Delta_g=\frac{2w}{G-1}$ is the width of one grid square.
Then, the defined grid consists of the grid points $(\gp x_m,\gp y_m)\in\grid^{w}_G(0,0),\,m\in\idx_{G^2}$.
\cref{fig:system_block_diagram} depicts the example grid $\grid^{w}_{5}(0,0)$.

\section{\loneloc} \label{sec:l1local}
Before solving the problem at hand, the statistical properties of the
\gls{rss} at sensors need to be studied.
\subsection{Sum of log normal random variables} \label{sec:sumLogNormalDist}
The sum of \gls{ln} random variables has an unknown \gls{pdf} \cite{fenton1960}, even for the sum of
two random variables. 
By using the Fenton-Wilkinson \cite{fenton1960} method, $r_k$ can be
approximated by an \gls{ln} random variable which has the same mean and variance as $r_k$.
Let $R_{kn}$ be a random variable from which the values of $r_{kn}$ are drawn, with
$r_{kn}$ being the \gls{rss} at sensor $k$ due to \ord n target, i.e.,
$r_{kn}=p_n\, d^{-\alpha}_{kn}\beta^{d_{kn}}\,10^\frac{\zeta_{kn}}{10}$. Then,
the mean and variance of $R_{kn}$ are given by \cite{zandi_wisee2018}:
\begin{align}
\EXPOP(R_{kn})=&p_n\, d^{-\alpha}_{kn}\beta^{d_{kn}}b_{kn}\,,\\
\Var(R_{kn})=&(p_n\, d^{-\alpha}_{kn}\beta^{d_{kn}})^2(b_{kn}^2-1)b_{kn}^2\,,
\end{align}
where $b_{kn}=e^{\frac{(\ln 10)^2\sigma^2_{kn}}{200}}$.
It is assumed that all $\sigma_{kn}$ are equal and $b=b_{kn}$ is known.
Since all the random variables $R_{kn}$ are pairwise independent the mean $M_k$ and variance $V_k$ of the random variable $R_k\coloneqq \sum\limits_{n\in\idx_N} R_{kn}$
reads
\begin{subequations}\label{eq:SLN_mean_var}
\begin{align}
M_k=&b g_k\,,\;g_k\coloneqq \sum\limits_{n\in\idx_N}p_n\, d^{-\alpha}_{kn}\beta^{d_{kn}}\,,\\
V_k=&(b^2-1)b^2 h_k\,,\;h_k\coloneqq  \sum\limits_{n\in\idx_N}(p_n\, d^{-\alpha}_{kn}\beta^{d_{kn}})^2\,.
\end{align}
\end{subequations}
The goal is to find $\muk$ and $\sigk$ such that the mean and the variance of the random variable $e^{\muk+\sigk X}$,
where $X$ is a standard normal random variable,
equate with the ones of $R_k$:
\begin{align}
M_k=&e^{\muk+\frac{\sigk^2}{2}}\,,\\
V_k=&e^{2\muk+2\sigk^2}\,,
\end{align}
By so doing the approximation
$\ln r_k \approx \muk+\sigk X$ is achieved.
This results in 
\begin{subequations}\label{eq:ln_rk_mean_var}
\begin{align}
\muk     &=2\ln(M_k)-\frac{1}{2}\ln(M_k^2+V_k)\,,\\
  \sigk^2&=\ln(M_k^2+V_k)-2\ln M_k\,.
\end{align}
\end{subequations}

\subsection{The General Idea of the Solution}
In this work, the equation \cref{eq:modeluncert_ple_basic_formula} is solved by a heuristic
to estimate the number of targets, their positions, and values of transmit power as
well as \gls{ple} and \gls{plf}. The heuristic is iterative and performs a sequence of actions at 
iteration $i$, given a grid and the estimation values from the previous 
iteration.
The concept of these actions is explained below:
\begin{enumerate}[(i)]
\item \defterm{Error minimization:} Minimizing an error function w.r.t to the variables
$\dxm$, $\dym$, $\dpm$, $\da$, $\db$, and $\nu$. The error function 
is a summation of two functions:
\begin{enumerate}[(1)]
\item The error function $\sum_{k\in\idx_K} {\linfk}^2$, where $\linfk$ is a linearization of the 
error function $f_k=\ln r_k-\gp \mu_k$, while $r_k$ is the actual \rss reading and
\begin{align}
\gp \mu_k&=\ln b+2\ln \gp g_k
-\frac{1}{2}\ln\left(\gp g_k^2+(b^2-1)\gp h_k\right)\,,\\
\gp g_k &= \sum\limits_{m\in\idx_M}\gp p_m\, \gp d^{-\alpha}_{km}\beta^{\gp d_{km}}\,,\\
\gp h_k &=  \sum\limits_{m\in\idx_M}(\gp p_m\, \gp d^{-\alpha}_{km}\beta^{\gp d_{km}})^2\,.
\end{align}

It needs to be mentioned that the accent \emph{tilde}
appearing over a variable means the area is discretized to $M$ grid points, while
the index $m$ refers to the \ord m grid point, for which the variable stands.

The (Taylor) linearization introduces the 
variables $\d \gp x_m$, $\d \gp y_m$, $\d \gp p_m$, $\d \alpha$, and $\d \beta$ which are
used for iterative update of the variables.
\item The error function $\normt{\bmr-\Phi\bms}^{^2}$, where $\bms$ is the selection vector and
$\obssym_{km}=[\sensmat]_{km}=\frac{1}{2}(\pmin+\pmax) d_{km}^{-\alpha}\beta^{d_{km}}$. The  \lzeronorm of the vector 
$\bms$ must be equal to $\nu$, i.e., must be $\nu$-sparse, such that only $\nu$ grid points are selected.
Note that $\gp s_m=1$ means
that \ord m grid point is chosen, while $\gp s_m=0$ means otherwise. 

This part of the objective function
does not include the transmit power, \gls{ple}, and \gls{plf} as optimization variables. It only selects the 
grid points such that the error is minimized, assuming the transmit power of each target is the average of $\pmin$ and $\pmax$.
\end{enumerate}
\item \defterm{\lone-relaxation:}Since an optimization that involves the \lzeronorm is \nphard, here this norm is relaxed
to \lonenorm, i.e., $s_m\in[0, 1]$, in order to convexify the underlying optimization.
On the other hand, due to the fact that $\sensmat$ does not hold incoherence properties,
the optimal vector $\opt\bms$ is not $N$-sparse. To over come this shortcoming, a \emph{cluster-and-average}
scheme is devised.
\item \defterm{Clustering:}
Let the set $\setSmax$ be the index set of $M_0$ largest entries of the optimal vector $\opt \bms$, 
given $M_0 < M$.
Then, the set of all positions $(\gp x_m+\d\gp x_m^\star,\gp y_m+\d\gp y_m^\star)$, $\forall m\in\setSmax$ is
represented by $\partSmax$.
Using the \kmean, $\partSmax$ can be
partitioned into $\parti_1,\cdots,\parti_N$, 
where $\parti_n\subset\partSmax\subset\grid(N,G)$, $\forall n\in\idx_N$.

\item\defterm{Averaging:} Then, the averaging rules
\begin{subequations}
\label{eq:l1local_averaging}
\begin{align}
  \est x_n &= \frac{\sum\limits_{m\in\parti_n} \opt s_m (\gp x_m^{i-1}+\d\gp x_m^\star)}{\sum\limits_{m\in\parti_n} \opt s_m}\,,
\label{eq:l1local_x_avg_I1}\\ 
  \est y_n &= \frac{\sum\limits_{m\in\parti_n} \opt s_m (\gp y_m^{i-1}+\d\gp y_m^\star)}{\sum\limits_{m\in\parti_n} \opt s_m}\,,
\label{eq:l1local_y_avg_I1}\\ 
  \est p_n &= \frac{\sum\limits_{m\in\parti_n} \opt s_m (\gp p_m^{i-1}+\d\gp p_m^\star)}{\sum\limits_{m\in\parti_n} \opt s_m}\,,
\label{eq:l1local_p_avg_I1}
\end{align}
\end{subequations}
are employed to update the position and power estimation.
\Cref{fig:l1local_averaging} shows how the cluster-and-average
improves positioning performance. Note that $M_0=N$ means that only
the $N$ largest entries of $\opt\bms$ and their corresponding grid points are selected
as the position estimation. This exclude averaging since each $\parti_n$ has
only one member.
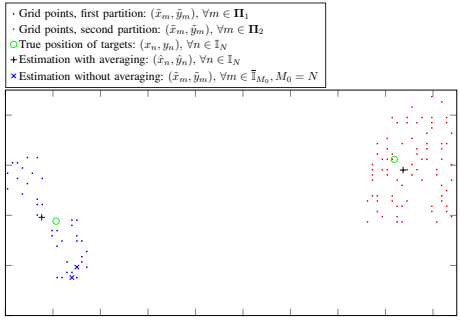
\begin{figure}
\begin{center}
  \scalebox{.6}{\def\markersize{2pt}
\begin{tikzpicture}
\begin{axis}[%
width=10cm,
height=5cm,
at={(1.733in,1.078in)},
scale only axis,
xmin=-800,
xmax=1100,
xticklabels={\empty},
ymin=-400,
ymax=500,
yticklabels={\empty},
axis background/.style={fill=white},
    legend style={nodes={scale=0.743, transform shape},at={(0.000,1.01)}, 
    anchor=south west, 
    legend cell align=left, 
    align=left, 
    draw=white!15!black}
]

\addplot [only marks, mark=*, mark options={solid, blue},mark size=0.2pt]
  table[row sep=crcr]{%
-499.234153080906	-248.419770865593\\
-540.900819747572	-227.586437532259\\
-561.734153080906	-248.419770865593\\
-811.734153080906	251.580229134407\\
-853.400819747572	147.413562467741\\
-811.734153080906	168.246895801074\\
-499.234153080906	-206.753104198926\\
-790.900819747572	168.246895801074\\
-540.900819747572	-185.919770865593\\
-770.067486414239	209.913562467741\\
-520.067486414239	-227.586437532259\\
-749.234153080906	209.913562467741\\
-832.567486414239	105.746895801074\\
-749.234153080906	147.413562467741\\
-811.734153080906	105.746895801074\\
-790.900819747572	105.746895801074\\
-582.567486414239	-248.419770865593\\
-728.400819747572	189.080229134407\\
-478.400819747572	-185.919770865593\\
-853.400819747572	84.9135624677406\\
-707.567486414239	126.580229134407\\
-728.400819747572	84.9135624677406\\
-561.734153080906	-102.586437532259\\
-478.400819747572	-165.086437532259\\
-707.567486414239	230.746895801074\\
-582.567486414239	-60.9197708655927\\
-686.734153080906	168.246895801074\\
-665.900819747572	64.0802291344073\\
-582.567486414239	-123.419770865593\\
-665.900819747572	43.2468958010739\\
-499.234153080906	-102.586437532259\\
-832.567486414239	64.0802291344073\\
-645.067486414239	43.2468958010739\\
-603.400819747572	-60.9197708655927\\
-457.567486414239	-206.753104198926\\
-499.234153080906	-81.7531041989261\\
-603.400819747572	-81.7531041989261\\
-520.067486414239	-40.0864375322594\\
-665.900819747572	1.58022913440725\\
-520.067486414239	-19.2531041989261\\
-478.400819747572	-102.586437532259\\
-853.400819747572	43.2468958010739\\
-645.067486414239	147.413562467741\\
-665.900819747572	230.746895801074\\
-499.234153080906	-19.2531041989261\\
-457.567486414239	-144.253104198926\\
};
\addlegendentry{Grid points, first partition: $(\gp x_m,\gp y_m),\,\forall m\in\parti_1$}

]

\addplot [only marks, mark=*, mark options={solid, red},mark size=0.2pt]
  table[row sep=crcr]{%
1015.88855188428	-26.1911389323961\\
1078.38855188428	-5.3578055990628\\
786.721885217616	182.142194400937\\
1057.55521855095	15.4755277342705\\
995.055218550949	15.4755277342705\\
1057.55521855095	36.3088610676039\\
995.055218550949	36.3088610676039\\
870.055218550949	-26.1911389323961\\
1036.72188521762	57.1421944009372\\
849.221885217616	-26.1911389323961\\
1036.72188521762	77.9755277342705\\
1057.55521855095	98.8088610676039\\
849.221885217616	-5.3578055990628\\
1036.72188521762	119.642194400937\\
870.055218550949	36.3088610676039\\
890.888551884282	57.1421944009372\\
1078.38855188428	140.475527734271\\
849.221885217616	36.3088610676039\\
1015.88855188428	161.308861067604\\
953.388551884282	140.475527734271\\
786.721885217616	-5.3578055990628\\
828.388551884282	57.1421944009372\\
1015.88855188428	182.142194400937\\
1078.38855188428	202.975527734271\\
828.388551884282	77.9755277342705\\
870.055218550949	119.642194400937\\
807.555218550949	182.142194400937\\
911.721885217616	161.308861067604\\
932.555218550949	182.142194400937\\
849.221885217616	140.475527734271\\
1036.72188521762	223.808861067604\\
765.888551884282	15.4755277342705\\
890.888551884282	182.142194400937\\
974.221885217616	223.808861067604\\
911.721885217616	202.975527734271\\
953.388551884282	223.808861067604\\
807.555218550949	161.308861067604\\
1036.72188521762	265.475527734271\\
932.555218550949	244.642194400937\\
765.888551884282	57.1421944009372\\
974.221885217616	265.475527734271\\
1057.55521855095	286.308861067604\\
1036.72188521762	286.308861067604\\
932.555218550949	265.475527734271\\
765.888551884282	77.9755277342705\\
974.221885217616	286.308861067604\\
828.388551884282	223.808861067604\\
1078.38855188428	327.975527734271\\
807.555218550949	223.808861067604\\
828.388551884282	244.642194400937\\
974.221885217616	327.975527734271\\
911.721885217616	307.142194400937\\
870.055218550949	286.308861067604\\
724.221885217616	-26.1911389323961\\
995.055218550949	348.808861067604\\
786.721885217616	223.808861067604\\
745.055218550949	77.9755277342705\\
1057.55521855095	369.642194400937\\
995.055218550949	369.642194400937\\
807.555218550949	265.475527734271\\
1057.55521855095	390.475527734271\\
932.555218550949	369.642194400937\\
995.055218550949	390.475527734271\\
745.055218550949	140.475527734271\\
953.388551884282	390.475527734271\\
745.055218550949	161.308861067604\\
890.888551884282	369.642194400937\\
745.055218550949	182.142194400937\\
765.888551884282	244.642194400937\\
807.555218550949	307.142194400937\\
890.888551884282	390.475527734271\\
745.055218550949	202.975527734271\\
849.221885217616	369.642194400937\\
724.221885217616	57.1421944009372\\
1036.72188521762	452.975527734271\\
911.721885217616	432.142194400937\\
765.888551884282	286.308861067604\\
828.388551884282	390.475527734271\\
995.055218550949	473.808861067604\\
};
\addlegendentry{Grid points, second partition: $(\gp x_m,\gp y_m),\,\forall m\in\parti_2$}

\addplot [only marks, mark=o, mark options={solid, green}, mark size=\markersize]
  table[row sep=crcr]{%
-586.561689321147	-23.1776224103417\\
837.221815875843	223.487725805291\\
};
\addlegendentry{True position of targets: $(x_n,y_n),\,\forall n\in\idx_N$}
\addplot [only marks, mark=+, mark options={solid, black},mark size=\markersize]
  table[row sep=crcr]{%
874.276135765096	180.766146573309\\
-647.419769282402	-7.95468405721618\\
};
\addlegendentry{Estimation with averaging: $(\est x_n,\est y_n),\,\forall n\in\idx_N$}

\addplot [only marks, mark=x, mark options={solid, blue},mark size=\markersize]
  table[row sep=crcr]{%
-499.234153080906	-206.753104198926\\
-520.067486414239	-248.419770865593\\
};
\addlegendentry{Estimation without averaging: $(\gp x_m,\gp y_m),\,\forall m\in\setSmax,M_0=N$}

\end{axis}
\end{tikzpicture}%
}
\caption{The result of localization with and without
cluster-and-average for $N=2$ targets.}
\label{fig:l1local_averaging}
\end{center}
\end{figure}
\item \defterm{Grid update:} Then, a sub-grid of $G^2$ points is formed around each estimate point and the power $\est p_n$ will
is associated with each point. This results in a set $\grid$ of $NG^2$ grid points
and the set $\power$ of their power values:
\begin{subequations}
\label{eq:l1local_grid_update}
\begin{align}
 \grid(N,G)&=\bigcup\limits_{n\in\idx_N} \grid^{\frac{w}{2}}_{G}(\est x_n,\est y_n)\,,\\
\power(N,G)&=\bigcup\limits_{n\in\idx_N} \est p_n\otimes \bOne_{G^2}\,,
\end{align}
\end{subequations}
where $2w$ is the area width, $\otimes$ is the Kronecker product, and $\bOne_{G^2}$ is the all-ones vector 
of size $G^2$. 
\item \defterm{Updating Number of Targets:} At each iteration, the number of targets is unknown and is represented by
the optimization variable $\nu\in\{\nmin,\cdots,\nmax\}$.
But to build the sets $\grid(N,G)$ and $\power(N,G)$, a value for $N$ is required.
Therefore, the optimal $\opt{\nu}$, from the previous iteration, is used.
It is reasonable to assume at the first iteration $N=\nmin$ target exist. The total number of grid points
is equal to $NG^2$ at each iteration.

\item \defterm{Updating $\alpha$ and $\beta$:} Those variables are updated using
\begin{subequations}
\label{eq:l1local_ab_update}
\begin{align}
     \alpha^i &=\alpha^{i-1}+\da^\star\,,\\
      \beta^i &=\beta^{i-1}+\db^\star\,.
\end{align}
\end{subequations}
\end{enumerate}

\subsection{Linearizing the Error Function $f_k$}
The first order Taylor series expansion is now deployed to linearize the error function
$f_k=\ln r_k-\gp \mu_k$.
The first derivative of $f_k$ w.r.t the variable $\theta$ is given by
\begin{align}
  \frac{\partial f_k}{\partial \theta}
  =-\frac{2}{\gp g_k}\frac{\partial \gp g_k}{\partial \theta}+\frac{\frac{\partial \gp g_k}{\partial \theta} \gp g_k
  +\frac{1}{2}(b^2-1)\frac{\partial \gp h_k}{\partial \theta}}{\gp g_k^2+(b^2-1)\gp h_k}\,,
\end{align}
where $\theta$ stands for $\gp x_m$, $\gp y_m$, $\gp p_m$, $\alpha$, and $\beta$. The derivatives
of the functions $\gp g_k$ and $\gp h_k$ read
\begin{subequations}
\begin{align}
\frac{\partial \gp g_k}{\partial \gp x_m}&=\frac{\gp p_m \,
\beta^{\gp d_{km}}\,(\sn x_k-\gp x_m)(\alpha-\gp d_{km}\,\ln \beta)} {\gp  d_{km}^{\alpha+2}}\,,\\
\frac{\partial \gp h_k}{\partial \gp x_m}&=2(\gp p_m\, \gp d^{-\alpha}_{km}\beta^{\gp d_{km}})\,
\frac{\partial \gp g_k}{\partial \gp x_m}\,,\\
\frac{\partial \gp g_k}{\partial \gp y_m}&=\frac{\gp p_m \,
\beta^{\gp d_{km}}\,(\sn y_k-\gp y_m)(\alpha-\gp d_{km}\,\ln \beta)} {\gp  d_{km}^{\alpha+2}}\,,\\
\frac{\partial \gp h_k}{\partial \gp y_m}&=2(\gp p_m\, \gp d^{-\alpha}_{km}\beta^{\gp d_{km}})\,
\frac{\partial \gp g_k}{\partial \gp y_m}\,,\\
\frac{\partial \gp g_k}{\partial \alpha}&=-\sum\limits_{m\in\idx_M}(\gp p_m\, \gp d^{-\alpha}_{km}\beta^{\gp d_{km}})\ln \gp d_{km}\,,\\
\frac{\partial \gp h_k}{\partial \alpha}&=-2\sum\limits_{m\in\idx_M}(\gp p_m\, \gp d^{-\alpha}_{km}\beta^{\gp d_{km}})^2\ln \gp d_{km}\,,\\
\frac{\partial \gp g_k}{\partial \beta}&=\sum\limits_{m\in\idx_M}(\gp p_m\, \gp d^{-\alpha}_{km}\beta^{\gp d_{km}})\gp d_{km}\beta^{-1}\,,\\
\frac{\partial \gp h_k}{\partial \beta}&=2\sum\limits_{m\in\idx_M}(\gp p_m\, \gp d^{-\alpha}_{km}\beta^{\gp d_{km}})^2\,\gp d_{km}\beta^{-1}\,,\\
\frac{\partial \gp g_k}{\partial \gp p_m}&= \gp d_{km}^{-\alpha}\beta^{\gp d_{km}}\,,\\
\frac{\partial \gp h_k}{\partial \gp p_m}&= 2\,\gp p_m\,\gp d_{km}^{-2\alpha}\beta^{2\gp d_{km}}\,.
\end{align}
\end{subequations}
Let 
{\small $[\gp p^{i-1}_1,\cdots,\gp p^{i-1}_M,\gp x^{i-1}_1,\cdots,\gp x^{i-1}_M\gp y^{i-1}_1,\cdots,\gp y^{i-1}_M,\alpha^{i-1},\beta^{i-1}]$}
be represented by the vector $\boldsymbol\theta$ at \ord i iteration,
then the coefficients $a^{i-1}_{km}$, $b^{i-1}_{km}$, $c^{i-1}_{km}$,
$u^{i-1}$, and $v^{i-1}$ stand for the derivatives of the error function $f_k$
w.r.t $\gp x_m$, $\gp y_m$, $\gp p_m$, $\alpha$, and $\beta$, respectively, at the point $\boldsymbol\theta$.
Then, $f_k$  at the \ord i iteration can be approximated by its first order Taylor term $\linfk$
\begin{align}
f_k&\approx \linfk=f^{i-1}_k +u^{i-1}\da+v^{i-1}\db+\nonumber\\
&\sum_{m\in\idx_M} a^{i-1}_{km}\d{\gp x}_m +b^{i-1}_{km}\d{\gp y}_m +c^{i-1}_{km}\d{\gp p}_m \,,
\end{align}
with $d \gp x_m$, $d \gp y_m$, $d \gp p_m$ , $\da$, and $\db$ being the optimization 
variables and $f_k^{i-1}\coloneqq f(\boldsymbol \theta)$.

\subsection{The Heuristic}
The proposed algorithm hinges upon solving the following convex \gls{qp} 
at each iteration, e.g., the \ord i iteration, for given $\grid(N,G)$, $\power(N,G)$, and $\mu\in\{0,1\}$:
\begin{subequations}\label{eq:modeluncert_problem_opt_combined}
	\begin{align}
\minimize{\substack{\nu, \d \alpha, \d \beta,\\s_m,\d{\gp x}_m,\d{\gp y}_m,\,\\\d{\gp p}_m\,, m\in\idx_M}		}\;
    \sum_{k\in\idx_K} \big[(f^{i-1}_k+u^{i-1}\da+v^{i-1}\db+\hspace{-0.2cm}\sum_{m\in\idx_M}\hspace{-0.2cm}a^{i-1}_{km}\d{\gp x}_m \nonumber\\
        +b^{i-1}_{km}\d{\gp y}_m+&c^{i-1}_{km}\d{\gp p}_m)^2  
   +\mu (\sum\limits_{k'\in\idx_K} \orthsym_{kk'}r_{k'}-\sum\limits_{m\in\idx_M}\cssym_{km}\,s_m)^2\big]
    \\
		\st s_m,\d{\gp x}_m,\,\d{\gp y}_m,\,\d{\gp p}_m,\,\nu\,,\da,\,\db\in\setR\,,\\
        & -\delta \leq \d\gp x_m \leq \delta\,,\\
        & -\delta \leq \d\gp y_m \leq \delta\,,\\
        &  \pmin-\gp p^{i-1}_m \leq \d\gp p_m \leq \pmax-\gp p^{i-1}_m \,,
        \label{eq:modeluncert_problem_opt_combined_power_const}\\
        &  \amin-\alpha^{i-1} \leq \da \leq \amax-\alpha^{i-1} \,,
        \label{eq:modeluncert_problem_opt_combined_alpha_const}\\
        & \bmin-\beta^{i-1} \leq \db \leq 1-\beta^{i-1} \,,
        \label{eq:modeluncert_problem_opt_combined_beta_const}\\
        & \nu\in\{\nmin,\cdots,\nmax\}\,,
        \label{eq:modeluncert_problem_opt_combined_N_const}\\
        & 0 \leq s_m \leq 1\,,
        \label{eq:modeluncert_problem_opt_combined_sm_range}\\
        & \sum\limits_{m\in\idx_M} s_m=\nu\,,
        \label{eq:modeluncert_problem_opt_combined_sm_sum}
	\end{align}
\end{subequations}
where $\orthsym_{kk'}$ and $\cssym_{km}$ are the entries of 
the pre-processing matrix $\orthmat$
and the matrix $\csmat=\orthmat\sensmat$ 
that is defined by
\begin{align}
\orthmat=\operatorname{orth}(\sensmat')'\sensmat^\dagger\,.
\end{align}
The symbol $^\dagger$ stands for the Moore–Penrose inverse and $\operatorname{orth}(\bX)$ is an orthogonal basis for the range of
matrix $\bX$. The authors in \cite{valaee2009} apply such a pre-processing by multiplying both sides of
the equation $\bmr=\sensmat\bms$ with $\orthmat$ since the sensing matrix $\sensmat$ does not
possess the \emph{incoherence} property.
The entry $km$ of the sensing matrix $\sensmat$ at \ord i iteration is given by 
\begin{align}
\obssym_{km}=\frac{\pmin+\pmax}{2}{(d^{i-1}_{km})}^{-\alpha^{i-1}}{(\beta^{i-1})}^{d^{i-1}_{km}}\,.
\end{align}

The constraint \crefrange{eq:modeluncert_problem_opt_combined_power_const}{eq:modeluncert_problem_opt_combined_N_const}
guarantee that the estimates $\est p_n$, $\est \alpha$, $\est \beta$, and $\est N$ are in their admissible ranges, given
the fact that the number of targets must be an integer.
Moreover, \cref{eq:modeluncert_problem_opt_combined_sm_range} together with 
\cref{eq:modeluncert_problem_opt_combined_sm_sum} relaxes
$\normz\bms$ to its \lonenorm.
Having incorporated the path loss model uncertainties as well as the number of targets,
\cref{alg:modeluncert} summarizes the \loneloc technique
to jointly estimate the number, position, and transmit power of the targets as well as
the values of \ple and \plf. 
Note that the estimation of $N$ can change from iteration to iteration.
After the iteration $I_1$ the variables $s_m$ become ineffective, since $\mu$ is set to zero.
Furthermore, the number of grid points becomes $N$ since $G$ becomes one. 

\begin{algorithm} 
  \caption{The \loneloc heuristic to jointly estimate the transmit power, positions of targets
  as well as the parameters of the path loss model in the presence of precipitation}
  \label{alg:modeluncert}
  \begin{algorithmic}  
    \State  {\bfseries initialization}:
    \begin{itemize}
      \item set the grid granularity $G\in\setN$
      \item set the area width $2w\geq 0$
      \item $\delta\leftarrow\frac{w}{4(G-1)}$
      \item $N\gets\nmin$
      \item let $\est p_n=\frac{1}{2}(\pmin+\pmax)$, $(\est x_n,\est y_n)=(0,0),\,\forall n\in\idx_N$
      \item let $\alpha^0=2$, $\beta^0=1$
      \item let $M_0=G^2$
      \item set the number of iterations by setting $I_1,I_2\in\setN$ 
      \item $\mu\gets 1$
    \end{itemize}
      \For{$i \gets 1$ to $I_1+I_2$}
      \If{$i =I_1+1$}
      \State $\mu\gets 0$
      \State $G\gets 1$
      \EndIf
        \State let $M=NG^2$ be the number of grid points
        \State define $\grid(\nu,G)$ and $\power(\nu,G)$ using \cref{eq:l1local_grid_update}
        \State let $(\gp x^{i-1}_m,\gp y^{i-1}_m)\in\grid(N,G)$ and $\gp p^{i-1}_m\in\power(N,G)$ 
        \State find $s_m^\star$, $\d\gp x_m^\star$, $\d\gp y_m^\star$, $\d\gp p_m^\star$, $\nu^\star$,
        $\da^\star$ and $\db^\star$ using \cref{eq:modeluncert_problem_opt_combined}
        \State find $\setSmax$ and $\parti_1,\cdots,\parti_{\nu^\star}$ using \kmean
        \State $N\gets\opt\nu$        
        \State update  $\est p_n$, $\est x_n$ and $\est y_n,\,\forall n\in\idx_{N}$
        using \cref{eq:l1local_averaging}
        \State calculate $\alpha^i$ and $\beta^i$ using \cref{eq:l1local_ab_update} 
        \EndFor 
    \State $\mathcal X\coloneqq\left\{ (\est x_n,\est y_n, \est p_n) \,|\,\forall n\in\idx_{\nu^\star}\right\}$
    \State $\mathcal A\coloneqq\left\{\est N, \est\alpha,\est\beta\,|\,\est N = \nu^\star, 
    \est\alpha=\alpha^{I_1+I_2}, \est\beta=\beta^{I_1+I_2}\right\}$
    \State {\bfseries return} $\mathcal X$ and $\mathcal A$
  \end{algorithmic}
\end{algorithm}
\subsection{Complexity Analysis} 

The complexity of the algorithm is governed by the \gls{qp}
\cref{eq:modeluncert_problem_opt_combined} whose complexity is $\bigo(n^3)$, where
$n$ is the number of variables.  In this problem, the number of variables itself varies
from iteration to iteration since
the number of targets is unknown. Nonetheless, the number of variables is upper-bounded by $4\nmax G^2+3$.
Furthermore, the complexity of the algorithm until the \ord{I_1} iteration overshadows the 
subsequent part of the algorithm since the number of grid points, i.e., the number of
optimization variables decreases afterward.
Note that the \kmean problem for the \emph{cluster-and-average}
can be solved by Lloyd's algorithm whose worst-case complexity is practically linear \cite{zandi_wsa2019}.
It can be then neglected since it is subordinate to the \gls{qp}.
Therefore, the upper bound on the complexity (the worst case) 
is $\bigo(64I_1\nmax^3G^6)$. Note that this upper bound is tight with $\nmax=N$ if the
number of targets is known. The complexity  analysis indicates that
estimating the \ple and \plf do not make the problem
much more complex.
        \begin{center}
            \fbox{\begin{minipage}{25em}
            \begin{itemize}[leftmargin=.3cm]
\item  From the viewpoint of complexity, it is worthwhile compensating for the path loss model uncertainties 
  by estimating $\alpha$ and $\beta$. 
\item From viewpoint of positioning error, it is crucial to estimate $\alpha$ and $\beta$. 
\item In case the number of targets is unknown, the localization fails without estimating $N$.
  \end{itemize}
            \end{minipage}}
        \end{center}

\section{Simulations} \label{sec:simulation}
\def\figsizeatsim{0.8}
Since there is no work with the same assumptions as this paper,
the results cannot be compared with any other works, unfortunately.
The only work, except for the previous papers \cite{zandi_iswcs2018, zandi_wisee2018} of the authors,
that has similar assumptions is \cite{jamali-rad_rss_2014}. It, nevertheless, deals with a fingerprinting 
problem. Therefore, a fair comparison with its results is not straightforward. 
In what follows the performance of the proposed \loneloc 
is evaluated by means of computer simulations.

In the simulation setup $\pmax=1$, $\pmin=0.5$ and $w=1 \text{Km}$ are chosen.
The results are the outcome of $J=5000$ simulation realizations, in each of which the
position of sensors and realization of $\zeta_{kn}$s are random, while
the transmit power and position of targets are always the same.
Let the estimated position of the \ord n target
 at \ord j realization be denoted by $(\est x^j_n,\est y^j_n)$. Then,
positioning \gls{rmse} in meters is defined by
\begin{equation} \label{eq:prmse}
\lrmse=\sqrt{\frac{1}{JN}\sum_{j=1}^{J}\sum\limits_{n=1}^N \left(\est x^j_n-x_n\right)^2+\left(\est y^j_n-y_n\right)^2}\ .
\end{equation}

Let the maximum positioning error at \ord j iteration, i.e.,
\begin{equation}
\lerr^j_\text{max}\coloneqq\max_{n\in\idx_N}\sqrt{\left(\est x^j_n-x_n\right)^2+\left(\est y^j_n-y_n\right)^2}\,,
\end{equation}
be a sample drawn from the distribution of a random variable, e.g., $\Delta$. Then, the error function 
\begin{equation}
P_{d}\coloneqq \Pr\left(\Delta > d\right)=1-F_{\Delta} (d)\,,
\end{equation}
stands for the probability that at least one of the targets is localized 
with an error of more than $d$ meters. Note that $F_\Delta$ is the empirical \gls{cdf} of the error $\Delta$.
Similarly, the \gls{rmse} of  the transmit power is defined by
\begin{equation}
\prmse=\sqrt{\frac{1}{NJ}\sum_{j=1}^{J}\sum_{n\in\idx_N} \perr^j_n}\,,
\end{equation}
where $\perr^j_n\coloneqq(p_n-\est p^j_n)^2$ is the square error of the estimated power 
of target $n$ at \ord j realization.
The \loneloc technique in \cref{alg:modeluncert} is simulated for three different 
scenarios:
\begin{enumerate}
\item The number of targets is known, but \ple and \plf are unknown, 
see \cref{fig:sim_modeluncert_cdf_N_2_alpha_unknown}.
The actual value of the \ple is chosen to be $\alpha=3.4$, 
while the actual value of \plf is $\beta=10^{-0.0015}\approx 0.99655$.
This value means the rain fade attenuation $\gamma_r=15$dB/km, e.g., 
the rainfall rate $R=100$mm/hr at 24GHz \cite[Fig. 7]{kestwal2014}. 
In the problem \cref{eq:modeluncert_problem_opt_combined} $\amin=1$,  $\amax=6$,
$\bmin=0.96$ and $\nmin=\nmax=N$.
\begin{figure}
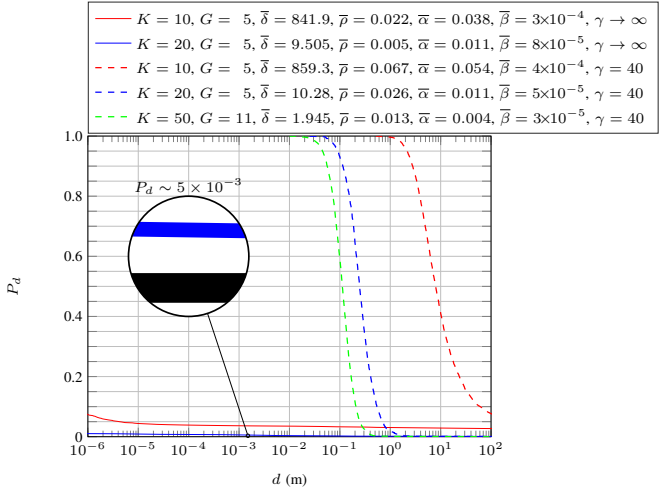

\begin{center}
\scalebox{\figsizeatsim}{\definecolor{iswcscolor}{rgb}{0.1,0.8,1.0}
\def\iswcslinewidth{0.8pt}
\begin{tikzpicture}[spy using outlines=
	{circle, magnification=35, connect spies}]
  \scriptsize
  \begin{axis}[%
    width=6.7cm,
    height=5cm,
    at={(1.106in,0.584in)},
    scale only axis,
    xmin=1e-6,
    xmax=1e2,
    xtick={1e-9, 1e-8, 1e-7, 1e-6, 1e-5, 1e-4, 1e-3, 1e-2, 1e-1, 1e0, 1e1, 1e2, 1e3},
    xlabel style={font=\color{white!15!black}},
    xlabel={$d$ (m)},
    xmode=log,
    ymin=0,
    ymax=1e0,
    ytick={0, 0.05, 0.1, 0.15, 0.2, 0.25, 0.3, 0.35, 0.4, 0.45, 0.5, 0.55, 0.6, 0.65, 0.7 ,0.75, 0.8, 0.85, 0.9, 0.95, 1.0},
    yticklabels={0, , , , 0.2, , , , 0.4, , , , 0.6, , , , 0.8, , , , 1.0},
    xmajorgrids,
    ymajorgrids,
    ytick style={font=\tiny \color{white!15!black}},
    ylabel style={yshift=0cm,font=\color{white!15!black}},
    ylabel={$P_{d}$},
    axis background/.style={fill=white},
    legend style={at={(0.000,1.008)}, 
    anchor=south west, 
    legend cell align=left, 
    align=left, 
    draw=white!15!black}
    ]
    \input{figures/mse_N_2_K_10_G_5_SNR_1000_I_20_Pratio_2_Nratio_1_cs.tex}
    \input{figures/mse_N_2_K_20_G_5_SNR_1000_I_20_Pratio_2_Nratio_1_cs.tex}

    \input{figures/mse_N_2_K_10_G_5_SNR_40_I_20_Pratio_2_Nratio_1_cs.tex}
    \input{figures/mse_N_2_K_20_G_5_SNR_40_I_20_Pratio_2_Nratio_1_cs.tex}
    \input{figures/mse_N_2_K_50_G_11_SNR_40_I_20_Pratio_2_Nratio_1_cs.tex}
    
    \coordinate (spypoint) at (axis cs:15e-4,3e-3);
    \coordinate (magnifyglass) at (axis cs:1e-4,0.6);

    \end{axis}

\spy [black, size=2.0cm] on (spypoint) in node[fill=white] at (magnifyglass);
 \draw (magnifyglass) node[anchor=south,xshift=0pt,yshift=27pt] {$P_d\sim 5\times 10^{-3}$};
\end{tikzpicture}%
}
  \caption{The error probability $P_d$ against positioning error $d$ for $N=2$ targets 
  achieved by \cref{alg:modeluncert}. The transmit power of the targets are unknown.
  The \ple and \plf are unknown and their actual values are $\alpha=3.4$ and $\beta\approx 0.99655$.
  The algorithm is deployed with $I_1=8$ and $I_2=12$ number of iterations.
  The values of $\lrmse$, $\prmse$, $\armse$ and $\brmse$ are shown in the legend.}
\label{fig:sim_modeluncert_cdf_N_2_alpha_unknown}
\end{center}
\end{figure}

\item The number of targets is unknown and can be between $\nmin=1$ and $\nmax=4$,
as shown in \cref{fig:sim_modeluncert_targnum_alpha_known}.
The values of $\alpha=2$ and $\beta=1$ are considered as known, which 
indicates that $\amin=\amax=2$ and $\bmin=1$.

\begin{figure}
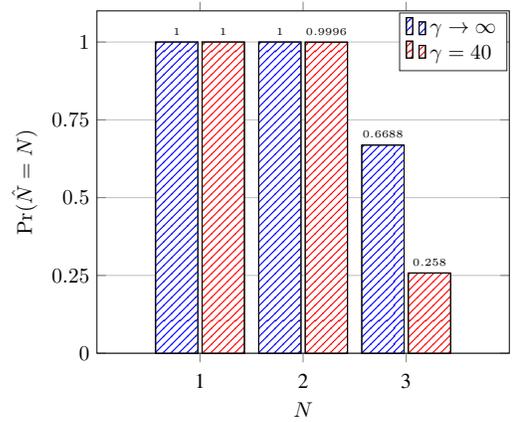
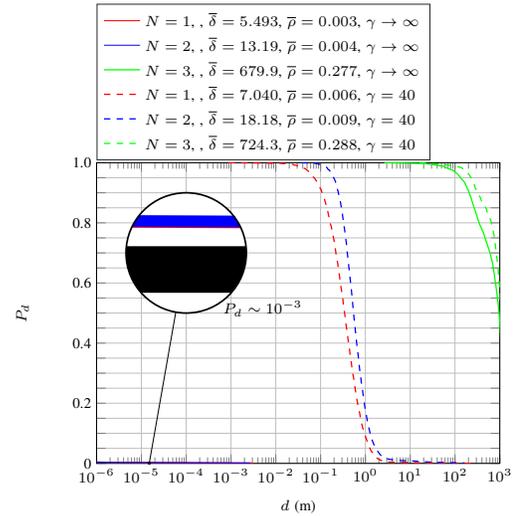

\begin{center}
\def\framesize@subfig{.46}
\def\subfigsize@local{\figsizeatsim}
\begin{subfigure}{\framesize@subfig\textwidth}
\begin{center}
\scalebox{\subfigsize@local}{\pgfplotsset{
  compat=newest,
  xlabel near ticks,
  ylabel near ticks
}
  \begin{tikzpicture}[
    ]
    \begin{axis}[
    ybar,
    bar width=20pt,
    xlabel={$N$},
    ylabel={$\Pr(\est N=N)$},
    ymin=0,
    ytick={0,0.25,0.50,0.75,1},
    xtick={1,2,3},
    ymajorgrids = true,
    enlarge x limits=0.5,
    symbolic x coords={1,2,3},
    xticklabel style={anchor=base,yshift=-\baselineskip},
    nodes near coords, 
    every node near coord/.append style={font=\tiny},
          /pgf/number format/precision=5,
    axis background/.style={fill=white},
    legend style={at={(0.995,0.995)}, 
    anchor=north east, 
    legend cell align=left, 
    align=left, 
    draw=white!15!black}
    ]
      \addplot[black,postaction={pattern=north east lines},pattern color=blue,fill=white] coordinates {
        (1,1)
        (2,1)
        (3,0.6688)
      };
    \addlegendentry{$\gamma\rightarrow\infty$}

      \addplot[black,postaction={pattern=north east lines},pattern color=red,fill=white] coordinates {
        (1,1)
        (2,0.9996)
        (3,0.258)
      };
    \addlegendentry{$\gamma=40$}
    \end{axis}
  \end{tikzpicture}}
\caption{The probability $\Pr(\est N=N)$ against number of targets $N$.}
\label{fig:sim_modeluncert_targnum_alpha_known_pmf}
\end{center}
\end{subfigure}
\vspace{0.5cm} \\
\begin{subfigure}{\framesize@subfig\textwidth}
\begin{center}
\scalebox{\subfigsize@local}{\definecolor{iswcscolor}{rgb}{0.1,0.8,1.0}
\def\iswcslinewidth{0.8pt}
\begin{tikzpicture}[spy using outlines=
	{circle, magnification=55, connect spies}]
  \scriptsize
  \begin{axis}[%
    width=6.7cm,
    height=5cm,
    at={(1.106in,0.584in)},
    scale only axis,
    xmin=1e-6,
    xmax=1e3,
    xtick={1e-9, 1e-8, 1e-7, 1e-6, 1e-5, 1e-4, 1e-3, 1e-2, 1e-1, 1e0, 1e1, 1e2, 1e3},
    xlabel style={font=\color{white!15!black}},
    xlabel={$d$ (m)},
    xmode=log,
    ymin=0,
    ymax=1e0,
    ytick={0, 0.05, 0.1, 0.15, 0.2, 0.25, 0.3, 0.35, 0.4, 0.45, 0.5, 0.55, 0.6, 0.65, 0.7 ,0.75, 0.8, 0.85, 0.9, 0.95, 1.0},
    yticklabels={0, , , , 0.2, , , , 0.4, , , , 0.6, , , , 0.8, , , , 1.0},
    xmajorgrids,
    ymajorgrids,
    ytick style={font=\tiny \color{white!15!black}},
    ylabel style={yshift=0cm,font=\color{white!15!black}},
    ylabel={$P_{d}$},
    axis background/.style={fill=white},
    legend style={at={(0.000,1.008)}, 
    anchor=south west, 
    legend cell align=left, 
    align=left, 
    draw=white!15!black}
    ]

    \input{figures/mse_N_1_K_20_G_5_SNR_1000_I_20_Pratio_2_Nratio_2_alpha_2_cs.tex}
    \input{figures/mse_N_2_K_20_G_5_SNR_1000_I_20_Pratio_2_Nratio_2_alpha_2_cs.tex}
    \input{figures/mse_N_3_K_20_G_5_SNR_1000_I_20_Pratio_2_Nratio_2_alpha_2_cs.tex}

    \input{figures/mse_N_1_K_20_G_5_SNR_40_I_20_Pratio_2_Nratio_2_alpha_2_cs.tex}
    \input{figures/mse_N_2_K_20_G_5_SNR_40_I_20_Pratio_2_Nratio_2_alpha_2_cs.tex}
    \input{figures/mse_N_3_K_20_G_5_SNR_40_I_20_Pratio_2_Nratio_2_alpha_2_cs.tex}

    \coordinate (spypoint) at (axis cs:15e-6,1e-3);
    \coordinate (magnifyglass) at (axis cs:1e-4,0.7);
    \coordinate (orderinfo) at (axis cs:1e-3,0.4);

    \end{axis}

\spy [black, size=2.0cm] on (spypoint) in node[fill=white] at (magnifyglass);
 \draw (magnifyglass) node[anchor=north west,xshift=15pt,yshift=-20pt] {$P_d\sim 10^{-3}$};
\end{tikzpicture}%
}
\caption{The error probability $P_d$ against positioning error $\d$.
The probability shows only those cases that the number of targets are correctly estimated.
The values of $\lrmse$ and $\prmse$ are shown in  the legend.}
\label{fig:sim_modeluncert_targnum_alpha_known_cdf}
\end{center}
\end{subfigure}
\caption{The localization performance  achieved by \cref{alg:modeluncert},
for an unknown number of targets.
The \ple and \plf are assumed to be known and $\alpha=2$ and $\beta=1$.
In the algorithm $N$ can be between $\nmin=1$ and $\nmax=4$.
The algorithm is deployed with $G=5$, $I_1=8$ and $I_2=12$ and the  number of
sensors is $K=20$.}
\label{fig:sim_modeluncert_targnum_alpha_known}
\end{center}
\end{figure}

\item Neither the number of targets, nor \ple nor \plf is known, 
see \cref{fig:sim_modeluncert_targnum_alpha_unknown}.
In this scenario, $\nmin=1$, $\nmax=4$ and the actual 
values of $\alpha$ and $\beta$ are chosen to be 3.4 and 0.99655, respectively.
In the constraint \cref{eq:modeluncert_problem_opt_combined_alpha_const} $\amin$=1 and $\amax=6$. 
In the constraint \cref{eq:modeluncert_problem_opt_combined_beta_const} also $\bmin$ is set to 0.96.
\begin{figure}
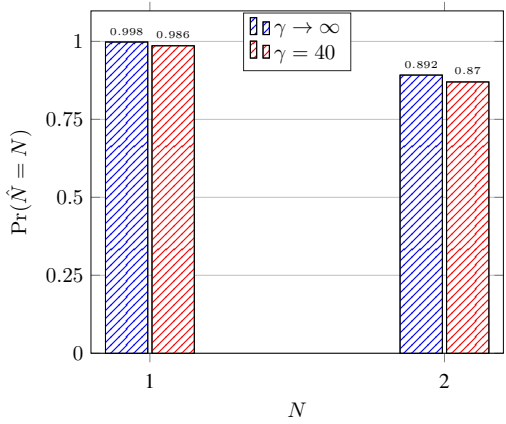
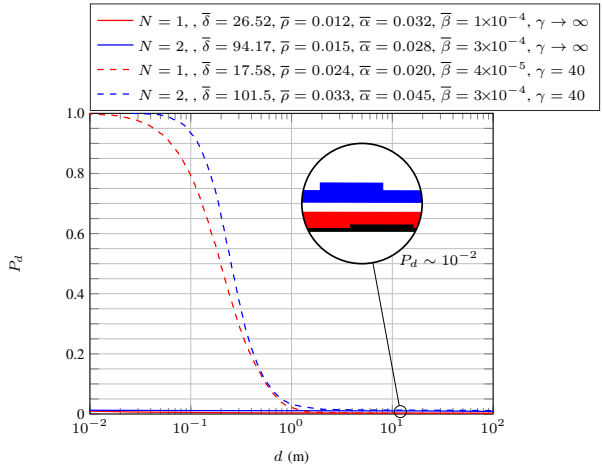

\begin{center}
\def\framesize@subfig{0.46}
\def\subfigsize@local{\figsizeatsim}
\begin{subfigure}{\framesize@subfig\textwidth}
\begin{center}
\scalebox{\subfigsize@local}{\pgfplotsset{
  compat=newest,
  xlabel near ticks,
  ylabel near ticks
}
  \begin{tikzpicture}[
    ]
    \begin{axis}[
    ybar,
    bar width=20pt,
    xlabel={$N$},
    ylabel={$\Pr(\est N=N)$},
    ymin=0,
    ytick={0,0.25,0.50,0.75,1},
    xtick={1,2,3},
    ymajorgrids = true,
    enlarge x limits=0.2,
    symbolic x coords={1,2,3},
    xticklabel style={anchor=base,yshift=-\baselineskip},
    nodes near coords, 
    every node near coord/.append style={font=\tiny},
          /pgf/number format/precision=5,
    axis background/.style={fill=white},
    legend style={at={(0.5,0.995)}, 
    anchor=north, 
    legend cell align=left, 
    align=left, 
    draw=white!15!black}
    ]
      \addplot[black,postaction={pattern=north east lines},pattern color=blue,fill=white] coordinates {
        (1,0.998)
        (2,0.892)
      };
    \addlegendentry{$\gamma\rightarrow\infty$}

      \addplot[black,postaction={pattern=north east lines},pattern color=red,fill=white] coordinates {
        (1,0.986)
        (2,0.87)
      };
    \addlegendentry{$\gamma=40$}
    \end{axis}
  \end{tikzpicture}}
\caption{The probability $\Pr(\est N=N)$ against number of targets $N$.}
\label{fig:sim_modeluncert_targnum_alpha_unknown_pmf}
\end{center}
\end{subfigure}
\vspace{0.5cm} \\
\begin{subfigure}{\framesize@subfig\textwidth}
\begin{center}
\scalebox{\subfigsize@local}{\definecolor{iswcscolor}{rgb}{0.1,0.8,1.0}
\def\iswcslinewidth{0.8pt}
\begin{tikzpicture}[spy using outlines=
	{circle, magnification=10, connect spies}]
  \scriptsize
  \begin{axis}[%
    width=6.7cm,
    height=5cm,
    at={(1.106in,0.584in)},
    scale only axis,
    xmin=1e-2,
    xmax=1e2,
    xtick={1e-9, 1e-8, 1e-7, 1e-6, 1e-5, 1e-4, 1e-3, 1e-2, 1e-1, 1e0, 1e1, 1e2, 1e3},
    xlabel style={font=\color{white!15!black}},
    xlabel={$d$ (m)},
    xmode=log,
    ymin=0,
    ymax=1e0,
    ytick={0, 0.05, 0.1, 0.15, 0.2, 0.25, 0.3, 0.35, 0.4, 0.45, 0.5, 0.55, 0.6, 0.65, 0.7 ,0.75, 0.8, 0.85, 0.9, 0.95, 1.0},
    yticklabels={0, , , , 0.2, , , , 0.4, , , , 0.6, , , , 0.8, , , , 1.0},
    xmajorgrids,
    ymajorgrids,
    ytick style={font=\tiny \color{white!15!black}},
    ylabel style={yshift=0cm,font=\color{white!15!black}},
    ylabel={$P_{d}$},
    axis background/.style={fill=white},
    legend style={at={(0.000,1.008)}, 
    anchor=south west, 
    legend cell align=left, 
    align=left, 
    draw=white!15!black}
    ]

    \input{figures/mse_N_1_K_20_G_5_SNR_1000_I_20_Pratio_2_Nratio_2_alpha_3.4_cs.tex}
    \input{figures/mse_N_2_K_20_G_5_SNR_1000_I_20_Pratio_2_Nratio_2_alpha_3.4_cs.tex}

     \input{figures/mse_N_1_K_20_G_5_SNR_40_I_20_Pratio_2_Nratio_2_alpha_3.4_cs.tex}
     \input{figures/mse_N_2_K_20_G_5_SNR_40_I_20_Pratio_2_Nratio_2_alpha_3.4_cs.tex}

    \coordinate (spypoint) at (axis cs:12e0,8e-3);
    \coordinate (magnifyglass) at (axis cs:5e0,0.7);

    \end{axis}

\spy [black, size=2.0cm] on (spypoint) in node[fill=white] at (magnifyglass);
 \draw (magnifyglass) node[anchor=north west,xshift=15pt,yshift=-20pt] {$P_d\sim 10^{-2}$};
\end{tikzpicture}%
}
\caption{The error probability $P_d$ against positioning error $\d$.
The probability shows only those cases that the number of targets are correctly estimated.
The values of $\lrmse$, $\prmse$, $\armse$ and $\brmse$ are shown in the legend.}
\label{fig:sim_modeluncert_targnum_alpha_unknown_cdf}
\end{center}
\end{subfigure}
\caption{The localization performance  achieved by \cref{alg:modeluncert},
for an unknown number of targets under path loss model uncertainties.
In the algorithm $N$ can be between $\nmin=1$ and $\nmin=4$.
The \ple and \plf are assumed unknown and their actual values are $\alpha=3.4$ and $\beta\approx 0.99655$.
The algorithm is deployed with $G=5$, $I_1=8$ and $I_2=12$ and the  number of
sensors is $K=20$.}
\label{fig:sim_modeluncert_targnum_alpha_unknown}
\end{center}
\end{figure}

\end{enumerate}
In all these three scenarios the transmit power of the targets are unknown.
Under the aforementioned assumptions, all algorithms in \cite{zandi_iswcs2018,zandi_wisee2018,zandi_wsa2019}
fail since uncertainties in values of $N$, $\alpha$, and $\beta$ are overlooked.

As it is evidenced by the \crefrange{fig:sim_modeluncert_cdf_N_2_alpha_unknown}{fig:sim_modeluncert_targnum_alpha_unknown}, the parameters of the
path loss model can be well estimated using the proposed algorithm.
For instance for $N=2$ and $\gamma\to\infty$, in case the number of targets is correctly estimated
{\small $\Pr(\Delta>1\text{mm})\approx 0$}. 
\Cref{fig:sim_modeluncert_targnum_alpha_known_pmf,fig:sim_modeluncert_targnum_alpha_unknown_pmf}
shows that for the case $N=1,2$, the probability
of the correct estimation of $N$ is between than 87\%-100\%. 
Obviously, for $N>2$ targets a much bigger number of sensors is required for a successful localization.

Furthermore, for $\gamma=40$ the positioning error of all the targets is very unlikely to be more than 10m, if the number of sensors is sufficient , i.e., $K>10$.  
In the strong shadowing conditions, i.e., higher values of $\sigma$, more \glspl{sn} can be deployed
to make the estimation more reliable. This is hopefully viable since \gls{rss}-based localization requires inexpensive and unsophisticated
sensors, on the one hand. On the other hand, the proposed \loneloc method has a low-complexity
and can solve the problem for higher values of $K$, efficiently.

\bibliographystyle{IEEEtran}
{\small
\bibliography{citations}

\begin{thebibliography}{10}
\providecommand{\url}[1]{#1}
\csname url@samestyle\endcsname
\providecommand{\newblock}{\relax}
\providecommand{\bibinfo}[2]{#2}
\providecommand{\BIBentrySTDinterwordspacing}{\spaceskip=0pt\relax}
\providecommand{\BIBentryALTinterwordstretchfactor}{4}
\providecommand{\BIBentryALTinterwordspacing}{\spaceskip=\fontdimen2\font plus
\BIBentryALTinterwordstretchfactor\fontdimen3\font minus
  \fontdimen4\font\relax}
\providecommand{\BIBforeignlanguage}[2]{{%
\expandafter\ifx\csname l@#1\endcsname\relax
\typeout{** WARNING: IEEEtran.bst: No hyphenation pattern has been}%
\typeout{** loaded for the language `#1'. Using the pattern for}%
\typeout{** the default language instead.}%
\else
\language=\csname l@#1\endcsname
\fi
#2}}
\providecommand{\BIBdecl}{\relax}
\BIBdecl

\bibitem{patwari2005}
N.~Patwari, J.~N. Ash, S.~Kyperountas, A.~O. Hero, R.~L. Moses, and N.~S.
  Correal, ``Locating the nodes: cooperative localization in wireless sensor
  networks,'' \emph{IEEE Signal Processing Magazine}, vol.~22, no.~4, pp.
  54--69, July 2005.

\bibitem{lee2009}
J.~H. Lee and R.~M. Buehrer, ``Location estimation using differential {RSS}
  with spatially correlated shadowing,'' in \emph{Global Telecommunications
  Conference, 2009. GLOBECOM 2009. IEEE}, 2009, pp. 1--6.

\bibitem{rappaport_book}
T.~Rappaport, \emph{Wireless Communications: Principles and Practice},
  2nd~ed.\hskip 1em plus 0.5em minus 0.4em\relax NJ, USA: Prentice Hall PTR,
  2001.

\bibitem{mao2007}
G.~Mao, B.~D.~O. Anderson, and B.~Fidan, ``Path loss exponent estimation for
  wireless sensor network localization,'' \emph{Computer Networks}, vol.~51,
  no.~10, pp. 2467--2483, Jul. 2007.

\bibitem{li_TVT_2007}
X.~Li, ``Collaborative localization with received-signal strength in wireless
  sensor networks,'' \emph{IEEE Transactions on Vehicular Technology}, vol.~56,
  no.~6, pp. 3807--3817, Nov 2007.

\bibitem{kong2008}
Q.~Kong, X.~Yang, and X.~Xie, ``A novel localization algorithm based on
  received signal strength ratio,'' in \emph{Wireless Communications,
  Networking and Mobile Computing, 2008. WiCOM'08. 4th International Conference
  on}.\hskip 1em plus 0.5em minus 0.4em\relax IEEE, 2008, pp. 1--6.

\bibitem{li_TWC_2006}
X.~Li, ``{RSS}-based location estimation with unknown pathloss model,''
  \emph{IEEE Transactions on Wireless Communications}, vol.~5, no.~12, pp.
  3626--3633, December 2006.

\bibitem{chan2012ccece}
Y.T., B.H., R., and F., ``Estimation of emitter power, location, and path loss
  exponent,'' in \emph{2012 25th IEEE Canadian Conference on Electrical and
  Computer Engineering (CCECE)}, April 2012, pp. 1--5.

\bibitem{hata80}
M.~Hata, ``\BIBforeignlanguage{English}{Empirical formula for propagation loss
  in land mobile radio services},'' \emph{\BIBforeignlanguage{English}{IEEE
  Transactions on Vehicular Technology}}, vol.~29, no.~3, pp. 317--325, 1980.

\bibitem{clark_book}
M.~P. Clark and M.~Clarke, \emph{Wireless access networks: fixed wireless
  access and WLL networks--design and operation}.\hskip 1em plus 0.5em minus
  0.4em\relax John Wiley \& Sons, Inc., 2000.

\bibitem{kestwal2014}
M.~C. Kestwal, , and S.~J. L.~S. Garia, ``Prediction of rain attenuation and
  impact of rain in wave propagation at microwave frequency for tropical region
  (uttarakhand, india),'' \emph{nternational Journal of Microwave Science and
  Technology}, pp. 1--6, 2014.

\bibitem{zandi_iswcs2018}
E.~Zandi and R.~Mathar, ``{RSS}-based location and transmit power estimation of
  multiple co-channel targets,'' in \emph{15th International Symposium on
  Wireless Communication Systems (ISWCS'18)}, Lisbon, Portugal, Aug. 2018, pp.
  1--6.

\bibitem{zandi_wisee2018}
E.~Zandi and R.~mathar, ``{RSS}-based positioning of multiple co-channel
  targets with unknown transmit power in a log-normal shadowing scenario,'' in
  \emph{2018 {IEEE} International Conference on Wireless for Space and Extreme
  Environments (WiSEE'18)}, Huntsville, AL, USA, Dec. 2018, pp. 1--6.

\bibitem{zandi_wsa2019}
E.~Zandi and R.~Mathar, ``Multi-node {RSS}-based localization with the aid of
  compressed sensing: An $\ell_1$-localization approach,'' in \emph{23rd
  International ITG Workshop on Smart Antennas (WSA 2019)}, Vienna, Austria,
  April 2019, pp. 1--8, to appear.

\bibitem{cardieri_rappaport2000}
P.~Cardieri and T.~S. Rappaport, ``Statistics of the sum of lognormal variables
  in wireless communications,'' in \emph{VTC2000-Spring. 2000 IEEE 51st
  Vehicular Technology Conference Proceedings (Cat. No.00CH37026)}, vol.~3, May
  2000, pp. 1823--1827 vol.3.

\bibitem{stuber_book}
G.~L. St\"{u}ber, \emph{Principles of Mobile Communication (2Nd Ed.)}.\hskip
  1em plus 0.5em minus 0.4em\relax Norwell, MA, USA: Kluwer Academic
  Publishers, 2001.

\bibitem{chan2016milcom}
F.~Chan, Y.~T. Chan, and R.~Inkol, ``Path loss exponent estimation and {RSS}
  localization using the linearizing variable constraint,'' in \emph{IEEE
  MILCOM 2016 - Military Communications Conference}, Nov 2016, pp. 225--229.

\bibitem{gerkmann2012}
T.~Gerkmann and R.~C. Hendriks, ``Unbiased {MMSE}-based noise power estimation
  with low complexity and low tracking delay,'' \emph{IEEE Transactions on
  Audio, Speech, and Language Processing}, vol.~20, no.~4, pp. 1383--1393, May
  2012.

\bibitem{beritelli2018}
F.~Beritelli, G.~Capizzi, G.~L. Sciuto, C.~Napoli, and F.~Scaglione, ``Rainfall
  estimation based on the intensity of the received signal in a lte/4g mobile
  terminal by using a probabilistic neural network,'' \emph{IEEE Access}, 2018.

\bibitem{fang2016}
S.~Fang and Y.~S. Yang, ``The impact of weather condition on radio-based
  distance estimation: A case study in gsm networks with mobile measurements,''
  \emph{IEEE Transactions on Vehicular Technology}, vol.~65, no.~8, pp.
  6444--6453, Aug 2016.

\bibitem{fang2018}
S.-H. Fang, Y.-C. Cheng, and Y.-R. Chien, ``Exploiting sensed radio strength
  and precipitation for improved distance estimation,'' \emph{IEEE Sensors
  Journal}, vol.~18, no.~16, pp. 6863--6873, 2018.

\bibitem{hosseini2010}
M.~Hosseini, H.~Chizari, C.~K. Soon, and R.~Budiarto, ``{RSS}-based distance
  measurement in underwater acoustic sensor networks: An application of the
  lambert w function,'' in \emph{2010 4th International Conference on Signal
  Processing and Communication Systems}, Dec 2010, pp. 1--4.

\bibitem{devasirvatham90}
D.~M.~J. Devasirvatham, C.~Banerjee, M.~J. Krain, and D.~A. Rappaport,
  ``Multi-frequency radiowave propagation measurements in the portable radio
  environment,'' in \emph{IEEE International Conference on Communications,
  Including Supercomm Technical Sessions}, vol.~4, April 1990, pp. 1334--1340.

\bibitem{fenton1960}
L.~Fenton, ``The sum of log-normal probability distributions in scatter
  transmission systems,'' \emph{IRE Transactions on Communications Systems},
  vol.~8, no.~1, pp. 57--67, March 1960.

\bibitem{valaee2009}
C.~Feng, S.~Valaee, and Z.~Tan, ``Multiple target localization using
  compressive sensing,'' in \emph{IEEE GLOBECOM 2009 - Global
  Telecommunications Conference}, Nov 2009, pp. 1--6.

\bibitem{jamali-rad_rss_2014}
H.~Jamali-Rad, H.~Ramezani, and G.~Leus, ``Sparsity-aware multi-source {RSS}
  localization,'' \emph{Signal Process.}, vol. 101, pp. 174--191, Aug. 2014.

\end{thebibliography}
}
\end{document}